\documentclass[a4paper]{jpconf}
\usepackage{graphicx}
\usepackage{cite}
\usepackage{amsmath}

\begin{document}
\title{CMB spectral distortions as a novel way to probe the small-scale structure problems}

\author{James A.~D.~Diacoumis and Yvonne Y.~Y.~Wong}

\address{School of Physics, The University of New South Wales, Sydney NSW 2052, Australia}

\ead{j.diacoumis@unsw.edu.au, yvonne.y.wong@unsw.edu.au}

\begin{abstract}
In this work we consider CMB spectral distortions as a probe of dark matter microphysics in the early universe. We demonstrate that future experiments such as PRISM have the potential to distinguish between scenarios which offer solutions to the small-scale problems of CDM cosmology. 
\end{abstract}

\section{Introduction}
Whilst the Cold Dark Matter (CDM) paradigm is remarkably successful at explaining the universe's large-scale structure there are many known issues pertaining to structure formation on small-length scales. These include the ``missing satellites'' \cite{MissingSat1, MissingSat2}, ``Too Big to Fail'' \cite{TBTF1, TBTF2} and ``core-cusp'' \cite{CuspCore1, CuspCore2, CuspCore3} problems all of which can broadly be thought of as an overprediction of density at small-scales by simulations of CDM when compared with observation. 

Consequently, solutions to these problems generally try to replicate the gross characteristics of CDM on large length scales where it performs well, whilst suppressing structure formation at small-scales where the CDM paradigm fares poorly. Examples of these solutions include Warm Dark Matter (WDM) \cite{WDM, WDM1, WDM2, WDM3} in which the dark matter is endowed with a velocity dispersion allowing it to stream freely and wash out density perturbations below some scale. Another example is Late Kinetic Decoupling (LKD) \cite{InteractingDM, InteractingDM2, LKD, LKD2, LKD3} in which the DM is allowed to couple to a relativistic ``heat bath'' (i.e. photons or standard model neutrinos) via elastic scattering until fairly late in the universe's evolution. The coupling allows small-scale modes (i.e. modes which entered the Hubble horizon early and therefore experience the strong effects of the coupling) to experience acoustic oscillations in their density while large scale modes enter the horizon after decoupling and therefore experience little or no suppression of power. 

Both of these solutions lead to, all other things considered, a sharp cutoff in the matter power spectrum and as a result can alleviate some of the problems for CDM at small-length scales. Given that these scenarios are identical at large length scales and possess the same phenomenological characteristics on the small, the question arises as to whether a cosmological observable exists which will allow for these scenarios to be distinguished. 

This manuscript is intended as a short summary of \cite{Diacoumis:2017hff}, in which we investigated CMB spectral distortions as a probe of these scenarios, and is organised as follows. In Section 2 we give a brief overview of CMB spectral distortions. In sections 3 and 4 we examine the details of DM--neutrino and DM--photon interactions respectively and comment on the effect that these would have on spectral distortions. In section 5 we discuss the implications of our results for future experiments such as PRISM. For further details the reader is encouraged to refer to \cite{Diacoumis:2017hff}.

\section{Spectral Distortions}

CMB Spectral Distortions are defined as any deviation in the energy spectrum of CMB photons from a perfect ``blackbody'' shape, they occur during periods where thermal equilibrium is not maintained in the photon fluid i.e. periods of significant energy injection or energy release. One of the most well-known mechanisms for producing CMB Spectral Distortions is through the dissipation of standing sound waves in the photon-baryon fluid whereby energy is transferred from the sound wave to the energy spectrum of the photons due to diffusion damping \cite{Hu:1993gc, ChlubaRev}. Since diffusion damping occurs at extremely small length scales \(1 \, \textrm{Mpc}^{-1} \, \leq k \leq 10^{4} \, \textrm{Mpc}^{-1}\),  roughly corresponding to the scales at which LKD and WDM offer solutions to the small-scale structure problems of CDM, we expect the details of the Dark Matter microphysics to have a significant impact on the form of the distortions.

We emphasise that spectral distortions are expected to occur via this mechanism even in CDM cosmology, the point is that DM microphysics can significantly alter the form of the perturbations which in turn leads to distortion patterns which deviate from CDM. We will refer throughout this manuscript to the \(\mu\)-parameter which characterises the ``effective chemical potential'' in the photon energy spectrum created as a result of the distortion.

\section{DM -- neutrino interactions}
In the absence of interactions with DM, neutrinos stream freely and therefore carry away some of the power of the perturbation without sourcing a spectral distortion. This is reflected in the form of the photon temperature transfer function which has the form 
\begin{equation} \label{eq:Theta}
\Theta_{1} \approx A \, c_{s} \sin\left(k r_{s}\right)\exp\left(-\frac{k^{2}}{k^{2}_{D}}\right).
\end{equation}
Here the acoustically oscillating part of the perturbation depends on the sound horizon \(r_{s}\), the diffusion damping is characterised by the diffusion scale \(k_{D}\), and the WKB amplitude is given by
\begin{equation} \label{eq:amp}
A \simeq \left(1 +\frac{4}{15}f_{\nu}\right)^{-1},
\end{equation}
where \(f_{\nu} = \rho_{\nu}/(\rho_{\nu} + \rho_{\gamma}) \simeq 0.41\) is the ratio of the neutrino energy density to the total energy density of the relativistic species. The factor of \(f_{\nu}\) amounts to a small correction to the amplitude as a result of the anisotropic stress of the neutrino fluid. 

A coupled DM--neutrino system exhibits similar characteristics to the familiar photon--baryon system in that the DM--neutrino fluid experiences acoustic oscillations damped by neutrino diffusion in the tight-coupling limit. The amplitude of the DM density perturbations is reduced on small length scales which serves as the basis for the solution to the small-scale problems of CDM cosmology. At the same time, coupling the neutrinos to DM prevents them from free-streaming and allows them to behave more like a `perfect' relativistic fluid with vanishing anisotropic stress \cite{GravCluster}.

\begin{figure}[h]
\includegraphics[width=22pc]{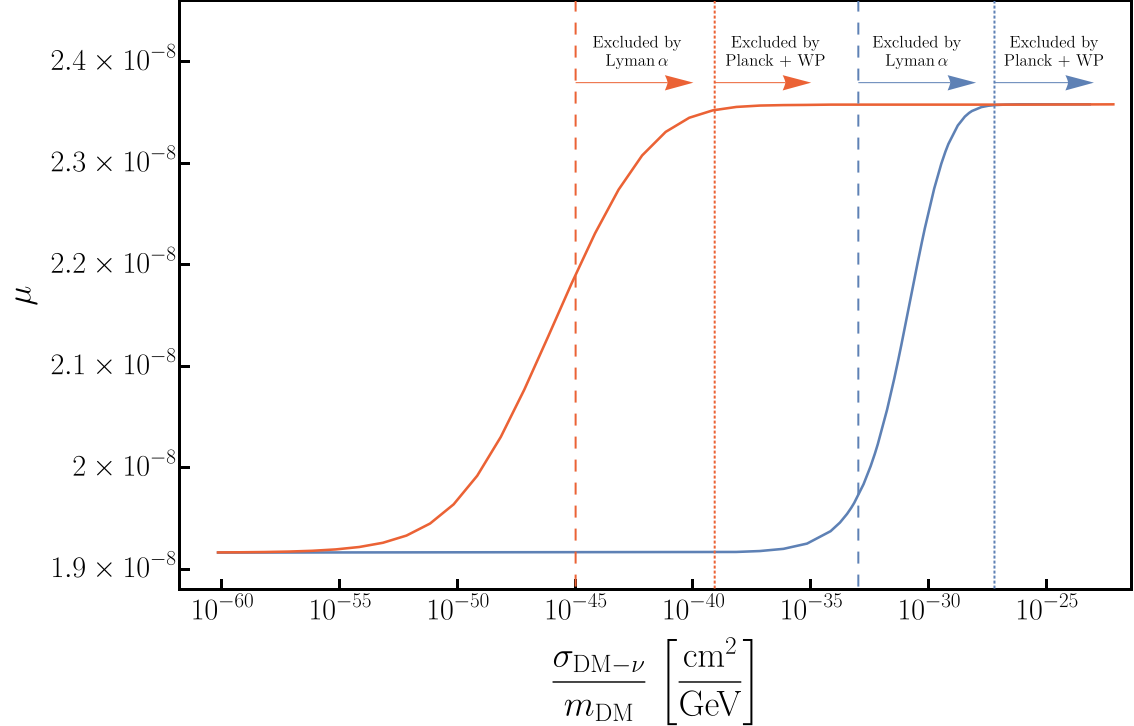}\hspace{2pc}%
\begin{minipage}[b]{10pc}\caption{\label{NeutrinoMu}Expected chemical potential \(\mu\)-distortion as a function of DM--neutrino scattering cross section \(\sigma_{DM-\nu}\) Blue denotes time independant cross-sections and red denotes cross sections proportional to temperature squared.\\ \\ \\}
\end{minipage}
\end{figure}

The absence of anisotropic stress allows the neutrinos to participate in acoustic oscillations of the perturbations in the same way as photons as they are no longer free-streaming. In the tightly coupled limit the correction factor \(f_{\nu}\) vanishes completely from the amplitude Eq. (\ref{eq:amp}) and the oscillation amplitude is enhanced relative to CDM. The net result is an \textit{enhanced} \(\mu\)-parameter as the amplitude of the perturbation has increased and there is more power that can potentially be dissipated leading to larger spectral distortions.
We emphasise that this effect is purely gravitational and arises because the neutrinos anisotropic stress vanishes in the presence of the interaction. Figure \ref{NeutrinoMu} shows clearly that this is a saturated effect and causes a maximum ~\(20\%\) increase in the \(\mu\)-parameter when the DM--neutrino coupling is tight.

\section{DM -- photon interactions} 
In the photon--baryon fluid the diffusion damping scale is set by 
\begin{equation} \label{eq:diffusion}
\partial_{z}k_{D}^{-2} = - \frac{c^{2}_{s}a}{2\mathcal{H}\dot{\kappa}}\left(\frac{16}{15} + \frac{R^{2}}{R+1}\right),
\end{equation}
where \(c_{s}\) is the sound speed of the photon--baryon fluid, \(\dot{\kappa}\) is the interaction rate of photons and baryons and \(R = (3/4)\rho_{b}/\rho_{\gamma}\) is the baryon-to-photon energy density ratio. Note that the first term in Eq. (\ref{eq:diffusion}) arises due to the viscosity of the photon fluid and the second term (proportional to \(R^{2} << 1\)) arises due to the heat conduction between photons and baryons, here the heat conduction term is extremely subdominant as the photons are tightly coupled to the baryons for the entire time period of interest and do not allow for efficient heat exchange between the two fluids \cite{Weinberg:1972kfs, Hu:1996mn}.

The case of DM--photon elastic scattering is distinctly non-trivial as the DM is not tightly coupled to the photons for the entirety of it's evolution and therefore enters a \textit{weak coupling regime} in which diffusion damping due to heat conduction can be significant or even the dominant mode of dissipation. Here the diffusion damping scale is given by
\begin{equation} \label{analytic}
\partial_{z}k_{D}^{-2} \simeq - \frac{c^{2}_{s}a}{2\mathcal{H}} \left[ \frac{1}{\dot{\kappa} + \dot{\mu}_{\gamma}}\frac{16}{15} + \frac{3\dot{\mu}}{k^{2}}\left(\frac{k^{2}}{k^{2} +3 S_{\gamma}^{-2}\dot{\mu_{\gamma}}^{2}}\right)  \right]
\end{equation}
where \(\dot{\mu}_{\gamma}\) is the DM--photon interaction rate and \(S = (3/4)\rho_{c}/\rho_{\gamma}\) is the DM-to-photon energy density ratio. When tight-coupling is satisfied (\(S^{-1}_{\gamma}\dot{\mu}_{\gamma} > k\)) the heat conduction term in this expression reduces to a form proportional to \(S_{\gamma}^{2}\) analogous to the heat conduction term in Eq. (\ref{eq:diffusion}). 

The effect of the heat conduction damping is twofold: firstly, it affects the evolution of the photon temperature transfer functions through additional damping by Eq. (\ref{eq:Theta}) which thereby damps the amount of perturbation power able to be dissipated and secondly, it directly affects the heating rate creating a temporally localised `burst' of energy. This second effect arises because the DM--photon coupling can enter a weak coupling regime during the \(\mu\)-era, in contrast with the photon--baryon coupling which remains tight for the entire duration of the \(\mu\)-era. To understand the `burst' shape consider that extremely tightly coupled fluids will be almost comoving and therefore not have a significant amount of heat conduction whereas fluid which are completely decoupled cannot communicate with each other efficiently enough to allow heat conduction to take place. The `burst' shape is therefore a result of the DM and photons slipping past each other during the weak couping regime and contributing to a large heat conduction between the two fluids. Both effects are demonstrated in Figure 2 which shows the heating rate of the CMB photons in which the `burst' shape at intermediate times and the suppression of power at late times are both visible.

\begin{figure}[h]
\begin{minipage}[c]{0.6\textwidth}
\includegraphics[width=22pc]{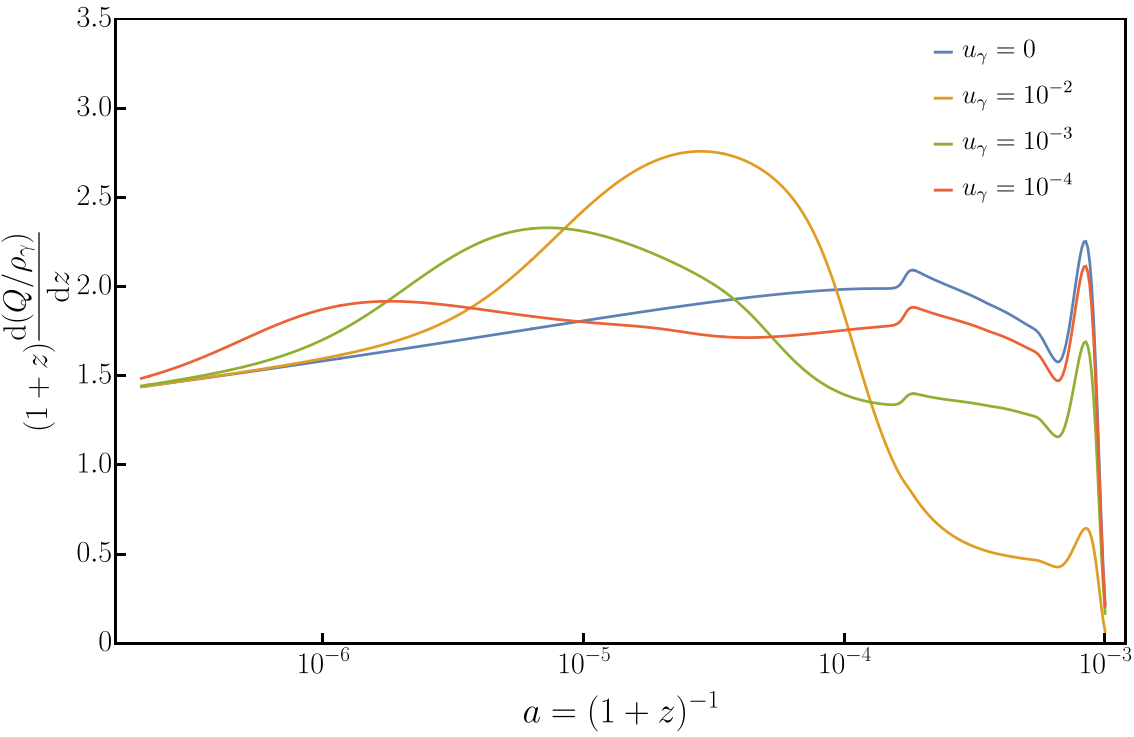}%
\end{minipage}\hspace{2pc}
\begin{minipage}[b]{0.3\textwidth}\caption{\label{PhotonHeatingConst}Heating rate of CMB photons as a function of the scale factor \(a\) for differerent values of the parameter \(u_{\gamma} \equiv \sigma_{\textrm{DM}-\gamma}/\sigma_{\textrm{T}}\left(100 \, \textrm{GeV}/m_{\textrm{DM}}\right)\). \\}
\end{minipage}
\end{figure}

\begin{figure}[h]
  \begin{minipage}[c]{0.6\textwidth}
    \includegraphics[width=22pc]{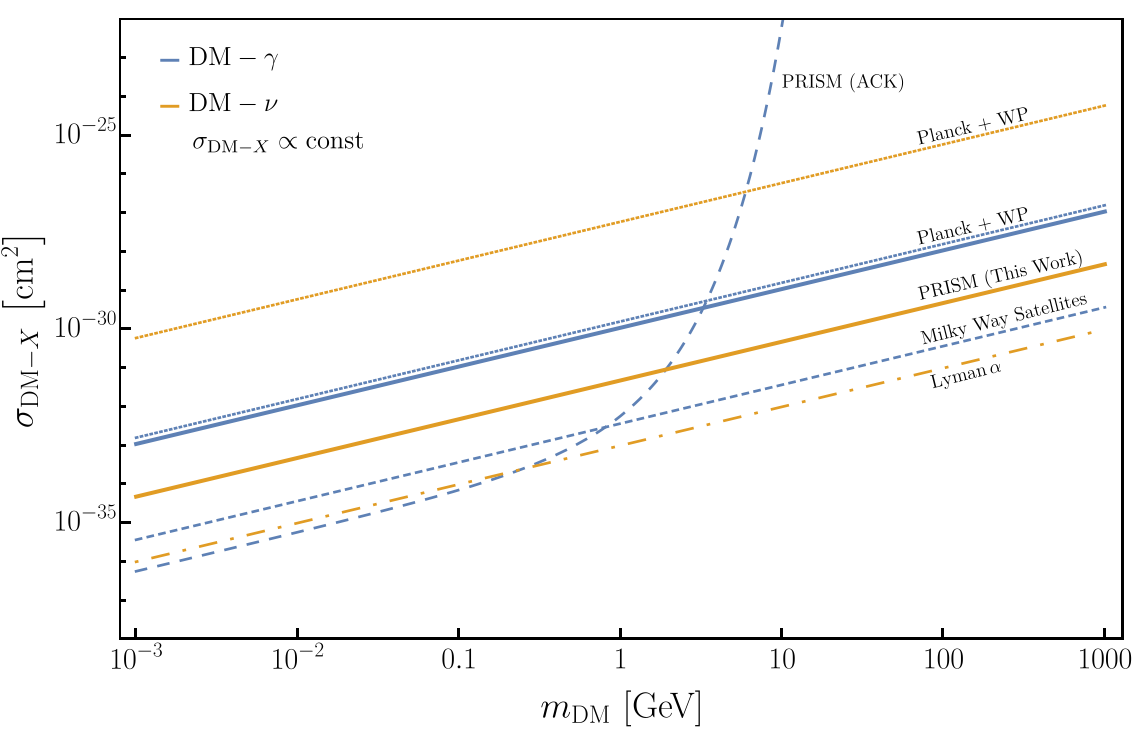}
    \includegraphics[width=22pc]{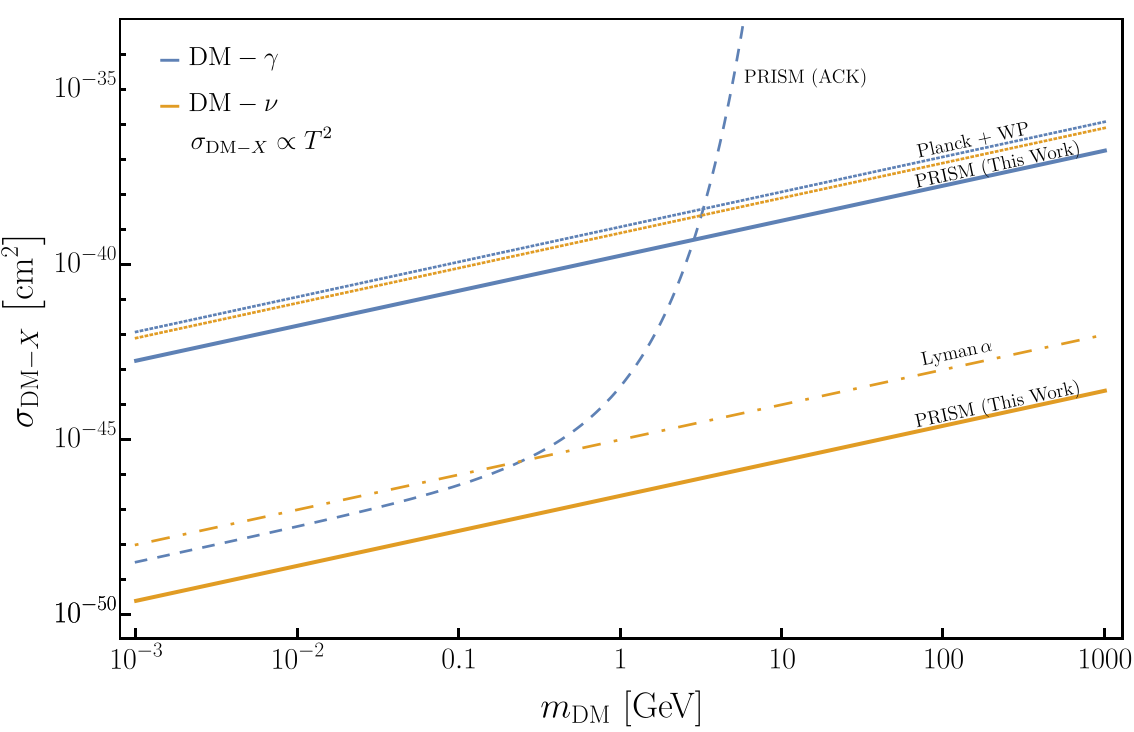}
  \end{minipage}\hspace{2pc}
  \begin{minipage}[c]{0.3\textwidth}
    \caption{\label{blah}\textit{Top}:  Upper bounds for DM-X scattering time independent cross-sections as a function of DM mass. Existing constraints from Planck + WP \cite{Wilknu, Wilkphoton}(dotted), MW satellites \cite{Boehm:2014vja} (small-dashed) and Lyman-\(\alpha\) forest \cite{Wilknu} (dot-dashed) shown alongside projected constraints from PRISM in \cite{ChlubaKam} (long-dashed) and this work (solid). Blue, orange denotes DM-\(\gamma\) and DM-\(\nu\) scattering respectively. \\ \\
       \textit{Bottom}: Same as top but for DM-X scattering cross-sections proportional to temperature squared as a function of DM mass. Existing constraints from Lyman-\(\alpha\) forest \cite{Wilknu} (dot-dashed) \\ \\ } \label{Constraints}
  \end{minipage}
\end{figure}

\section{Results}
In the case of DM--neutrino elastic scattering the overall amplitude of the perturbations is enhanced by \(\sim 10\%\) relative to \(\Lambda\)CDM in the limit of tight coupling. As a result, the expected \(\mu\) distortion is enhanced significantly (see Figure \ref{NeutrinoMu}) and may be distinguished from the \(\Lambda\)CDM predicition if the present-day value of the scattering cross section is at least as large as \(\sigma_{\textrm{DM}-\nu} \sim 4.8 \times 10^{-32}  \left(m_{\textrm{DM}}/\textrm{GeV}\right) \, \textrm{cm}^{2}\) for time-independent cross sections, and \(\sigma^{0}_{\textrm{DM}-\nu} \sim 2.5 \times 10^{-47}  \left(m_{\textrm{DM}}/\textrm{GeV}\right) \, \textrm{cm}^{2}\) for $\sigma_{{\rm DM}-\gamma} \propto T^2$.   In the latter case, it is interesting to note that the constraining power of PRISM on dark matter--neutrino elastic scattering may potentially exceed current limits from the Lyman-\(\alpha\) forest (see bottom of Figure \ref{Constraints}).

In the case of DM--photon elastic scattering we derive a new analytical expression Eq. (\ref{analytic}) for the diffusion damping scale in the presence of a DM--photon elastic scattering interaction which acccounts for the fact that the DM--photon fluid enters a weak coupling regime while the photons are still tightly coupled to the baryons. This expression shows that dissipation due to heat conduction is significant in the case of DM--photon elastic scattering and can actually be the dominant mode of dissipation during the \(\mu\)-era for certain parameter values. We find that future experiments such as PRISM may be sensitive to DM--photon elastic scattering if the cross section is at least as large as \(\sigma_{\textrm{DM}-\gamma} \sim 1.1 \times 10^{-30}  \left(m_{\textrm{DM}}/\textrm{GeV}\right) \, \textrm{cm}^{2}\) for time-independent cross sections, and \(\sigma^{0}_{\textrm{DM}-\nu} \sim 1.8 \times 10^{-40}  \left(m_{\textrm{DM}}/\textrm{GeV}\right) \, \textrm{cm}^{2}\) for $\sigma_{{\rm DM}-\gamma} \propto T^2$. This method is not as constraining as the case of DM--neutrino elastic scattering due to competing effects dominating the heating rate of the photons at different times (see Section 3 for further discussion.)

In summary, we have identified a number of physical features of DM--neutrino and DM--photon interacting models which influence the evolution of photon perturbations in the early universe. These physical features produce different CMB Spectral Distortions and can be probed with a high degree of sensitivity by future experiments such as PRISM.

\section*{References}
\bibliography{ref}

\bibliographystyle{iopart-num}

\end{document}